\documentclass[11pt]{article}
\usepackage{moriond,epsfig,amsmath,amssymb}
\usepackage{feynmf}
\usepackage{subfigure}
\usepackage{psfrag}

\newcommand{\be}{\begin{equation}}  
\newcommand{\ee}{\end{equation}}

\def\be{\begin{equation}}
\def\ee{\end{equation}}
\def\bea{\begin{eqnarray}}
\def\eea{\end{eqnarray}}

\def\slash#1{#1\!\!\!/\!\,\,}

\begin{document}
\begin{fmffile}{feynmf_file}
\vspace*{4cm}
\title{SOME NEW IMPLICATIONS OF THE ANOMALOUS BARYON CURRENT IN THE
STANDARD MODEL}

\author{RICHARD. J. HILL}

\address{Fermi National Accelerator Laboratory \\
P.O. Box 500, Batavia, Illinois 60510, USA
}

\maketitle\abstracts{
Phenomenological implications of the anomalous baryon current
in the Standard Model are discussed, in particular neutrino-photon
interactions at finite baryon density. 
A pedagogical derivation of the baryon current anomaly is
given.  
}

\section{Introduction}
The baryon current in the Standard Model
is not conserved in the presence of electroweak gauge fields.  
Although classically we have
\be
\partial_\mu J^\mu = \partial_\mu \left( 
\frac13 \sum_q \bar{q}\gamma^\mu q
\right)
 = 0 \,,
\ee
the baryon current divergence acquires quantum corrections
when gauge fields 
are coupled differently to left- and right-handed quarks.
For the Standard Model electroweak gauge fields, we have~\cite{'t Hooft:1976up}
\be\label{eq:bar_div}
\partial_\mu J^\mu = -{1\over 64\pi^2} \epsilon^{\mu\nu\rho\sigma}
\left( 
g_2^2 F^a_{\mu\nu}F^a_{\rho\sigma} 
- 
g_1^2 F^Y_{\mu\nu}F^Y_{\rho\sigma} 
\right) 
\ne 0 \,,
\ee
where $F^a_{\mu\nu} = \partial_\mu W^a_\mu - \partial_\nu W^a_\mu 
+g_2 \epsilon^{abc} W^b_\mu W^c_\nu$ is the 
covariant $SU(2)_L$ field strength and 
$F^Y_{\mu\nu}$ is the weak hypercharge field strength. 
This curious fact may have profound cosmological implications
through the generation of baryon number at the electroweak phase
transition~\cite{Kuzmin:1985mm}.

As discussed in Refs.~\cite{Harvey:2007rd,Harvey:2007ca} 
and reviewed in this talk, nonconservation of baryon number is connected 
to novel effects that can be observed in laboratory experiments, and 
that may have interesting astrophysical implications.  
This report begins with a theoretical review by analyzing the 
baryon number anomaly in analogy to the perhaps more familiar 
axial anomaly. 
Turning to phenomenology, some observable consequences in neutrino scattering
experiments are described, and several other directions to 
explore are mentioned. 

\section{Theoretical excursion} 

\subsection{The axial current anomaly and $\pi^0\to\gamma\gamma$}

A famous implication of gauge anomalies is the 
necessity for a nonzero $\pi^0\to\gamma\gamma$ amplitude 
due to the nonconservation 
of the iso-triplet axial-vector quark current,
\be
J^{5\mu} = \frac12( 
\bar{u} \gamma^\mu \gamma_5 u 
- 
\bar{d} \gamma^\mu \gamma_5 d 
)
\,.
\ee 
In the presence of electromagnetism we have~\cite{BJ,Adler}
\be\label{eq:axialdiv}
\partial_\mu J^{5\mu} 
 = {e^2 \over 32\pi^2} \epsilon^{\mu\nu\rho\sigma} F_{\mu\nu}F_{\rho\sigma} \,.
\ee
If low-energy QCD is described by a theory of mesons, 
a nonzero $\pi^0\to\gamma\gamma$ amplitude is necessary in order to 
reproduce this result. 

Let us recall how this works explicitly, by considering the object: 
\vspace{4mm}
\be
\int d^4x\, e^{-iq\cdot x} \langle\gamma(p)\gamma(k)|J^{5\mu}(x)|0\rangle
\equiv 
\Bigg[
\hspace{15mm} 
\parbox{30mm}{
\begin{fmfgraph*}(40,40)
\fmfleft{l}
\fmfrightn{r}{2} 
\fmf{photon}{l,v}
\fmf{photon}{v,r1}
\fmf{photon}{v,r2} 
\fmfblob{.36w}{v} 
\fmflabel{$B_5(q,\mu)$}{l}
\fmflabel{$A(p,\nu)$}{r1}
\fmflabel{$A(k,\rho)$}{r2}
\end{fmfgraph*}
}
\hspace{-5mm} 
\Bigg]
\times \epsilon^*_\nu(p) \epsilon^*_\rho(k) (2\pi)^4\delta^4(p+k-q) \,,
\vspace{3mm}
\ee
first at the quark level, and then at the meson level.  
The field $B_5$ denotes a background field coupled to $J^{5\mu}$, 
and $A$ is the photon. 
At the quark level, after a proper definition of 
the relevant triangle diagram that ensures vector current conservation,
a standard calculation~\cite{Peskin:1995ev} 
shows that in lowest order perturbation theory, 
\vspace{4mm}
\be\label{eq:diagramdiv}
iq_\mu
\Bigg[ 
\hspace{10mm} 
\parbox{30mm}{
\begin{fmfgraph*}(40,40)
\fmfleft{l}
\fmfrightn{r}{2} 
\fmf{photon}{l,v}
\fmf{photon}{v,r1}
\fmf{photon}{v,r2} 
\fmfblob{.36w}{v} 
\fmflabel{$B_5$}{l}
\fmflabel{$A$}{r1}
\fmflabel{$A$}{r2}
\end{fmfgraph*}
}
\hspace{-10mm}
\Bigg]
= 
{e^2\over 4\pi^2} \epsilon^{\nu\rho\alpha\beta} 
p_\alpha k_\beta \,, 
\vspace{3mm}
\ee
consistent with (\ref{eq:axialdiv}).  
How is this result reproduced in terms of the low-energy effective
action where the quarks are replaced by mesons?   
First, there is no gauge invariant operator connecting $B_5$ and two 
photons directly, so that 
\vspace{4mm}
\be
\parbox{30mm}{
\begin{fmfgraph*}(40,40)
\fmfleft{l}
\fmfrightn{r}{2} 
\fmf{photon}{l,v}
\fmf{photon}{v,r1}
\fmf{photon}{v,r2} 
\fmfdot{v} 
\fmflabel{$B_5$}{l}
\fmflabel{$A$}{r1}
\fmflabel{$A$}{r2}
\end{fmfgraph*}
}
\hspace{-10mm}
=0 \,.
\vspace{3mm}
\ee
A nonzero contribution is however obtained from the pion pole
(consider the limit of vanishing quark masses), 
\vspace{5mm}
\be\label{eq:piondiv}
\parbox{30mm}{
\begin{fmfgraph*}(40,40)
\fmfleft{l}
\fmfrightn{r}{2} 
\fmf{phantom}{l,v}
\fmf{photon}{v,r1}
\fmf{photon}{v,r2}
\fmffreeze
\fmf{double, label=$\pi$}{v,w}
\fmf{photon}{w,l} 
\fmfdot{v} 
\fmfdot{w}
\fmflabel{$B_5$}{l}
\fmflabel{$A$}{r1}
\fmflabel{$A$}{r2}
\end{fmfgraph*}
}
\hspace{-10mm}
= 
{- i C_1 q^\mu} 
\times
{i\over q^2} 
\times 
(-i C_2) {e^2\over 4\pi^2}  \epsilon^{\nu\rho\alpha \beta} 
p_\alpha k_\beta \,.
\vspace{3mm}
\ee
Here $C_1$ denotes the strength of the
$\pi$ coupling to the axial current, and 
$C_2$ is the strength 
of the pion-photon vertex.  
From the chiral lagrangian with Wess-Zumino-Witten term~\cite{Wess,Witten}, 
we necessarily have $C_1= f_\pi$, $C_2=1/f_\pi$. 
Contracting (\ref{eq:piondiv}) 
with $iq_\mu$ reproduces (\ref{eq:diagramdiv}) and hence 
(\ref{eq:axialdiv}). 
Phrased differently, if low-energy QCD is described by
an effective theory of pions, then the process $\pi^0\to\gamma\gamma$
occurs with a fixed strength. 

\subsection{The baryon current anomaly}

The anomalous baryon current can be treated in close 
analogy to the anomalous axial-vector current above.  
We must however pay close attention to which currents are 
conserved, since in the present case it is no longer true that
``vector currents are conserved, axial-vector currents are anomalous,''
as the usual intuition suggests.
Suppose that we introduce a background field $B_\mu$
coupled to baryon number.  
Then the baryon current is defined by varying the action
with respect to $B_\mu$:
\be
J^\mu = {\delta S \over \delta B_\mu} \,, 
\ee
and its divergence is read off from 
\be\label{eq:Svar}
\delta S = -\int d^4x\, \epsilon(x)\, \partial_\mu J^\mu \,,
\ee
where $\delta B_\mu = \partial_\mu \epsilon$.   
Thus the problem of calculating the anomalous divergence
of the baryon current 
is reduced to the introduction of $B_\mu$. 

However, we must not be too naive in introducing $B_\mu$; 
otherwise we may start with a gauge invariant theory, but end up with 
a non-gauge-invariant (i.e., nonsensical) theory.   
Varying the nonsensical theory would not give the correct symmetry
current and its divergence.  
To be explicit, let us return to the example of 
the axial-vector current for a single fermion, 
and suppose that we add the perturbation
\be\label{eq:pert}
\bar{\psi}(i\slash{\partial} + \slash{A} )\psi 
\to 
\bar{\psi}(i\slash{\partial} + \slash{A} + \slash{B}_5\gamma_5  )\psi  \,. 
\ee
Then the theory naively remains invariant under electromagnetic 
gauge transformations, 
\be
\psi\to e^{i\epsilon} \psi\,,
\quad 
A_\mu \to A_\mu + \partial_\mu \epsilon \,,
\quad
B_{5\mu} \to B_{5\mu} \,.
\ee
However, due to the effects of anomalies~%
\footnote{That is, due to 
the effects of the fermion measure, in path integral language.}
the theory is in fact not gauge invariant.   
For a sensible theory, 
we must add at the same time as the perturbation (\ref{eq:pert}), 
a counterterm: 
\be
\bar{\psi}(i\slash{\partial} + \slash{A} )\psi 
\to 
\bar{\psi}(i\slash{\partial} + \slash{A} + \slash{B}_5\gamma_5  )\psi 
+ {\cal L}_{\rm ct}(A,B_5) \,,
\ee
where explicitly, 
\be\label{eq:bc}
{\cal L}_{\rm ct}(A,B_5) = {1\over 6\pi^2}\epsilon^{\mu\nu\rho\sigma}B_\mu A_\nu \partial_\rho A_\sigma \,. 
\ee
The results (\ref{eq:axialdiv}) and (\ref{eq:diagramdiv}) 
have an implicit dependence on the choice of counterterm.~%
\footnote{
The dependence can be made explicit by
performing the calculation with Weyl fermions.  
}  
In particular, the ``Bardeen''~\cite{Bardeen} 
form of the counterterm, of which (\ref{eq:bc}) is an example, 
is employed to conserve vector currents in the
presence of arbitrary background fields.   

When nonvectorlike currents are gauged, a different counterterm
must be employed.  For the general case the explicit counterterm is 
given in Ref.~\cite{Harvey:2007ca}.   
Let us consider the baryon current for a single 
standard model generation, and for simplicity restrict attention
to the neutral gauge bosons $A$ and $Z$.   The Bardeen 
counterterm is then
\be\label{eq:bardeen}
{\cal L}_{\rm Bardeen} 
= {eg_2\over 24\pi^2 \cos\theta_W} \epsilon^{\mu\nu\rho\sigma} 
(B_\mu Z_\nu \partial_\rho A_\sigma + A_\mu Z_\nu \partial_\rho B_\sigma )
\,,
\ee
whereas the full counterterm is 
\be\label{eq:ct}
{\cal L}_{\rm ct} 
= {eg_2\over 24\pi^2 \cos\theta_W} \epsilon^{\mu\nu\rho\sigma}
( -2B_\mu Z_\nu \partial_\rho A_\sigma + A_\mu Z_\nu \partial_\rho B_\sigma )\,.
\ee
If we now write 
\be\label{eq:Scomb}
S = [S + S_{\rm Bardeen} - S_{\rm ct} ] + S_{\rm ct} - S_{\rm Bardeen} \,,
\ee
then the variation (\ref{eq:Svar}) 
vanishes for the bracketed combination 
in (\ref{eq:Scomb}), and from the remainder we can read off 
immediately using (\ref{eq:bardeen}) and (\ref{eq:ct}):
\be
\partial_\mu J^\mu = -{eg_2\over 8\pi^2 \cos\theta_W} \epsilon^{\mu\nu\rho\sigma} 
\partial_\mu A_\nu \partial_\rho Z_\sigma \,,
\ee
yielding the result (\ref{eq:bar_div}).  

In the language of chiral lagrangians, the 
new counterterm has the novel property that it leaves residual  
``pseudo-Chern Simons'' terms in the action, i.e., terms 
involving the epsilon tensor, but having no pion fields. 
Such terms are subtracted if the Bardeen counterterm is 
used instead, since it can be shown that 
${\cal L}_{\rm Bardeen} = -{\cal L}(\pi=0)$. 
Equivalently, the counterterm requires a different 
boundary condition for ``integrating the anomaly'' to obtain
the anomalous part of the chiral lagrangian~\cite{Wess}; this 
is again related to the statement that ${\cal L}(\pi=0) \ne 0$.  

\subsection{Vector mesons} 

The preceding discussion shows how to 
incorporate spin-1 background fields into the 
chiral lagrangian without upsetting gauge invariance in
the fundamental gauge fields.  
In particular, the $SU(2)_L\times U(1)_Y$ 
gauge anomaly cancellation between quark and lepton 
sectors is not upset when background fields are coupled
to the quark flavor symmetries.  With these background 
field probes in place, it is straightforward to derive 
properly defined (covariant) currents and the associated
anomalous divergences, by an appropriate variation of 
the action.

The relevance of the background field discussion for
vector mesons is twofold.  First, 
since physical spin-1 mesons such as $\rho$ and $\omega$ 
behave mathematically like these background fields, 
we have found the ``slots'' which these fields fit into when
constructing our chiral lagrangian.  
Second, and relatedly, once we know that the vector mesons
inhabit these slots, we find new physical effects related
to the quark level anomalies, e.g. to the baryon current anomaly, 
These effects can be observed experimentally.  
For example, at the level of vector meson dominance, new effects
will be described by the interaction 
\be\label{eq:newint}
{\cal L} = {eg_2 \over 8\pi^2\cos\theta_W} \epsilon^{\mu\nu\rho\sigma}\omega_\mu Z_\nu \partial_\rho A_\sigma \,.
\ee
This is in the same spirit as using $\pi^0\to\gamma\gamma$ as a probe
of the axial current anomaly.  

The theoretical description can be refined at low energy by 
integrating out the vector mesons; the vector dominance assumption
then translates into a prediction for the coefficients of certain 
$1/m_\omega^2$ suppressed operators.

\section{Phenomenology}

\begin{figure}
\begin{center}
\begin{fmfchar*}(40,45)
\fmftopn{t}{2}
\fmfbottomn{b}{1}
\fmf{double}{b1,v}
\fmf{fermion}{t1,v,t2}
\fmflabel{$N$}{t1}
\fmflabel{$N$}{t2}
\fmfdot{v}
\fmfblob{4mm}{b1}
\end{fmfchar*}
\hspace{15mm}
\begin{fmfgraph*}(40,40)
\fmftopn{t}{3}
\fmfbottomn{b}{1}
\fmf{double}{b1,v}
\fmf{double}{t1,v}
\fmf{double}{t2,v}
\fmf{double}{t3,v}
\fmflabel{$\pi$}{t1}
\fmflabel{$\pi$}{t2}
\fmflabel{$\pi$}{t3}
\fmfdot{v}
\fmfblob{4mm}{b1}
\end{fmfgraph*}
\hspace{15mm}
\begin{fmfgraph*}(40,40)
\fmftopn{t}{2}
\fmfbottomn{b}{1}
\fmf{double}{b1,v}
\fmf{double}{t1,v}
\fmf{photon}{t2,v}
\fmflabel{$\pi$}{t1}
\fmflabel{$\gamma$}{t2}
\fmfdot{v}
\fmfblob{4mm}{b1}
\end{fmfgraph*}
\hspace{15mm}
\begin{fmfgraph*}(40,40)
\fmftopn{t}{2}
\fmfbottomn{b}{1}
\fmf{double}{b1,v}
\fmf{zigzag}{t1,v}
\fmf{photon}{t2,v}
\fmflabel{$Z$}{t1}
\fmflabel{$\gamma$}{t2}
\fmfdot{v}
\fmfblob{4mm}{b1}
\end{fmfgraph*}
\end{center}
\vspace{-2mm} 
\caption{\label{fig:barcur}
Different parts of the baryon current.  The bottom leg 
denotes a field such as $\omega$ coupling to this current, 
and the blob denotes a source of baryon number, such as a nucleus.}
\end{figure}
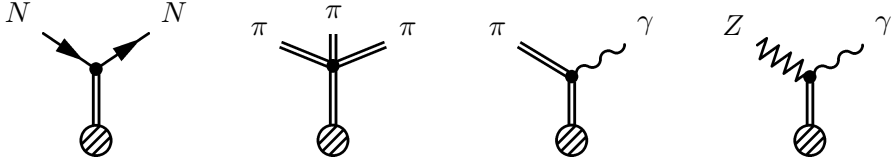
To see that the vector mesons are indeed described as part of the
WZW term structure, we can verify that the same coupling 
strength is observed in accessible decay modes, 
such as $\omega \to 3\pi$ and $\omega\to\pi\gamma$.  
As depicted in the Fig.~\ref{fig:barcur}, these are all parts of the 
baryon current, expressed in terms of the fields, including
nuclear sources, in the low-energy chiral lagrangian:
\be
J^\mu = \bar{N}\gamma^\mu N + {1\over 4\pi^2} \epsilon^{\mu\nu\rho\sigma}
\left( -{2i\over f_\pi^3} 
\partial_\nu\pi^+ \partial_\rho\pi^- \partial_\sigma\pi^0 
- {e\over f_\pi} 
\partial_\nu \pi^0 \partial_\rho A_\sigma
+ {eg_2 \over 2\cos\theta_W} Z_\nu \partial_\rho A_\sigma
+ \dots  
\right) \,. 
\ee
For example, 
\begin{align}
\Gamma(\omega\to\pi\gamma) \approx 
{3\alpha E_\gamma^3 \over 64\pi^4 f_\pi^2}\left(\frac23 g_\omega\right)^2
 \approx 0.76\,{\rm MeV} \left( \frac23 g_\omega\over 6 \right)^2 \,.
\end{align} 
Similarly, $\frac23 g_\omega \approx 6$ is obtained for $\omega\to3\pi$, including
the $\omega\to \rho\pi \to 3\pi$ contributions.~%
\footnote{
Note that $g_\omega=\frac32 g'$ in the conventions of
Ref.~\cite{Harvey:2007rd,Harvey:2007ca}. 
}
A consistent, although somewhat uncertain, 
value of the $\omega$ coupling to the baryon current 
is also obtained for the first diagram in Fig.~\ref{fig:barcur}, 
using one-meson exchange models of the 
force between nucleons, and isolating the isoscalar $J^P=1^-$ 
channel~\cite{Machleidt:1987hj}.  
The effective coupling is expected to be somewhat larger
in this case, since ``$\omega$'' is actually 
representing a tower of resonances. 

\vspace{3mm}
\begin{figure}[h]
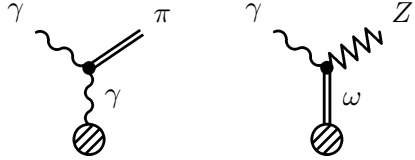

\begin{center}
\begin{fmfchar*}(40,45)
\fmftopn{t}{2}
\fmfbottomn{b}{1}
\fmf{photon,label=$\gamma$}{b1,v}
\fmf{photon}{t1,v}
\fmf{double}{v,t2}
\fmflabel{$\gamma$}{t1}
\fmflabel{$\pi$}{t2}
\fmfdot{v}
\fmfblob{4mm}{b1}
\end{fmfchar*}
\hspace{15mm}
\begin{fmfchar*}(40,45)
\fmftopn{t}{2}
\fmfbottomn{b}{1}
\fmf{double,label=$\omega$}{b1,v}
\fmf{photon}{t1,v}
\fmf{zigzag}{v,t2}
\fmflabel{$\gamma$}{t1}
\fmflabel{$Z$}{t2}
\fmfdot{v}
\fmfblob{4mm}{b1}
\end{fmfchar*}
\end{center}
\vspace{-2mm}
\caption{\label{fig:electric}
Analogy to the Primakoff effect: on the left, one of the
photons in the $\pi\gamma\gamma$ vertex couples to electric charge; 
on the right, $\omega$ from the $\omega Z\gamma$ vertex couples 
to baryon number.}
\end{figure}

We wish to access the final diagram in Fig.~\ref{fig:barcur}, 
i.e., the pure gauge field part of the baryon current, 
that is most directly related to the baryon current anomaly.  
We expect $\omega$ to couple to this part of the current 
with the same strength as the other parts.  
Now, if the $Z$ mass were small,~%
\footnote{
Consider the limit $m_{u,d}\to 0$, $g_2\to 0$ with $v$ fixed, so
that $m_\pi^2 \ll m_Z^2 \ll m_\omega^2$.  Then the $Z$ will eat 
mostly Higgs field, and effects of $\pi - Z$ mixing can be ignored. 
} the Standard Model 
would predict a decay mode, 
\be
\Gamma(\omega\to Z\gamma) = {3\alpha\over 256\pi^4} 
{E_\gamma^3\over m_Z^2} {g_2^2\over \cos^2\theta_W}\left(\frac23 g_\omega\right)^2
\left(1+{m_Z^2\over m_\omega^2} \right) \,.
\ee
Of course, the decay $\omega \to Z \gamma$ is not physically 
allowed.
Nevertheless, processes involving virtual $Z^*$ are allowed, and can lead 
to interesting effects.
Since there will be a weak suppression, we should focus on situations 
in which the $Z$ is ``useful'', e.g. processes involving neutrinos 
or parity violation.  We can also make the $\omega$ ``useful'', e.g., 
by utilizing its strong coupling to baryon number to look 
for enhanced rates when scattering off nuclei, rather than searching
for the tiny branching fraction $\omega\to\gamma\nu\bar{\nu}$.  
As depicted in Fig~\ref{fig:electric}, 
this is in analogy to probing the $\pi^0\gamma\gamma$ 
coupling via the Primakoff effect, where one of the photons couples 
coherently to the electric charge of the nucleus. 

\subsection{Neutrino scattering} 

\begin{figure}
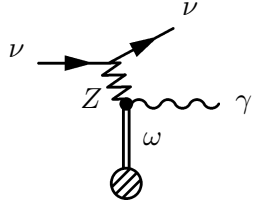

\begin{center}
\begin{fmfchar*}(70,60)
\fmftopn{t}{5}
\fmfbottomn{b}{5}
\fmfleftn{l}{5}
\fmfrightn{r}{5}
\fmf{phantom}{b3,v,t3}
\fmf{phantom}{l3,v,r3}
\fmffreeze
\fmf{phantom}{l4,w1,w2,w3,w4,r4}
\fmffreeze
\fmf{double,label=$\omega$}{b3,v}
\fmf{fermion}{l4,w2,t4}
\fmfblob{4mm}{b3}
\fmf{zigzag, label=$Z$, label.side=right}{w2,v}
\fmf{photon}{v,r3}
\fmflabel{$\nu$}{l4}
\fmflabel{$\nu$}{t4}
\fmflabel{$\gamma$}{r3}
\fmfdot{v}
\end{fmfchar*}
\end{center}
\vspace{-2mm}
\caption{\label{fig:nuscat}
Photon production in neutrino scattering in the presence of 
baryon number.
}
\end{figure}

The interaction (\ref{eq:newint}) will induce neutrino-photon interactions
in the presence of baryon number.  
For example, single photons are produced in neutrino-nucleus scattering, 
as depicted in Fig.~\ref{fig:nuscat}. 
In the approximation where the nuclear interactions are described by 
one-meson exchange, there will be competing contributions from 
virtual $\pi^0$ and $\rho^0$ exchange.  However, $\pi^0$ exchange 
is suppressed by the accidental smallness of $1-4\sin^2\theta_W$ 
in the Standard Model, and the $\rho^0$ exchange diagram is
suppressed in amplitude by $\sim (1+1-1)^2/(1+1+1)^2 = 1/9$ relative 
to $\omega$, due to the fact that $\omega$ is isoscalar, whereas 
$\omega$ is isotriplet; this can be thought of as a coherence 
effect at the nucleon level.  Further enhancement of the $\omega$ 
exchange due to coherence over adjacent nucleons can occur
in kinematics where small enough momentum is exchanged with the 
nucleus.  

\vspace{-2mm} 
\begin{figure}[h]
\begin{center}
\subfigure{
\psfrag{x}{$E_\gamma(\rm GeV)$}
\psfrag{y}{$d\sigma/dE_\gamma$}
\includegraphics[width=.45\textwidth]{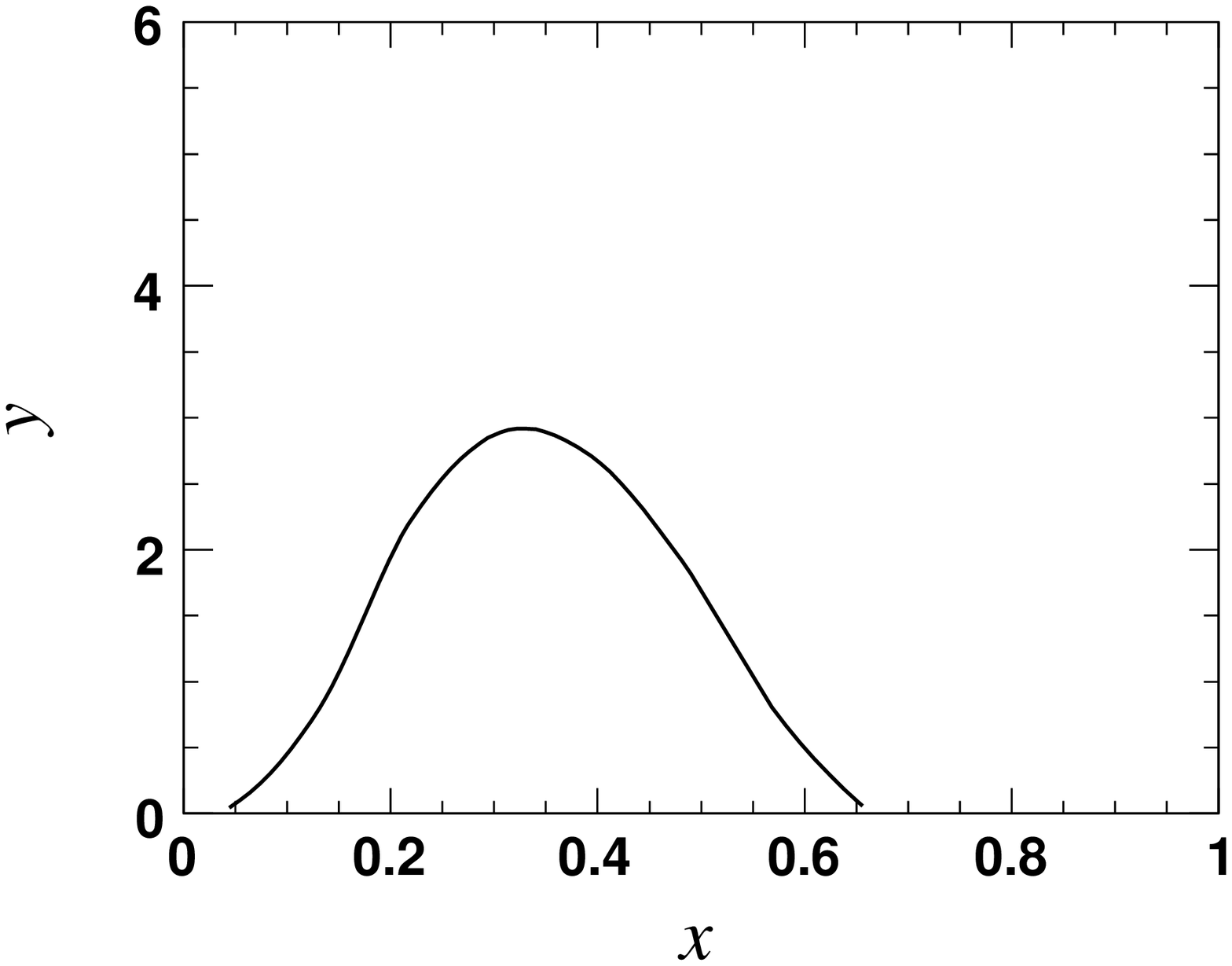}
}
\subfigure{
\psfrag{x}{$\cos\theta_\gamma$}
\psfrag{y}{$d\sigma/d\cos\theta_\gamma$}
\includegraphics[width=.45\textwidth]{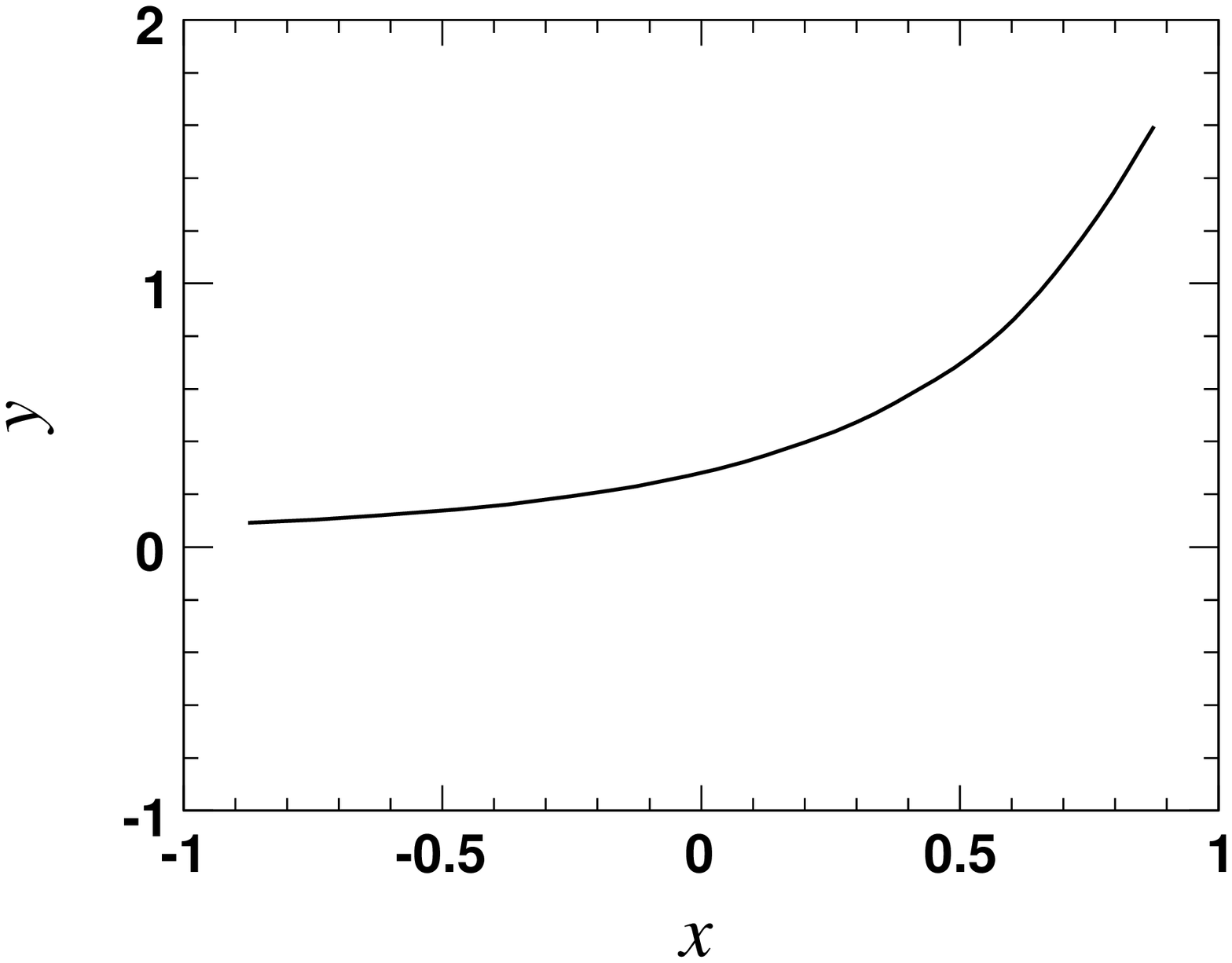}
}
\end{center}
\vspace{-7mm}
\caption{\label{fig:ff}
Photon energy distribution (left figure) and angular distribution 
(right figure), including nuclear recoil and $\omega(780)$ form 
factor, for $700\,{\rm MeV}$ neutrino incident on stationary nucleon
(arbitrary normalization).
}
\end{figure}

Neglecting effects such as coherence, form factors and recoil, the 
cross section for the process depicted in Fig.~\ref{fig:nuscat} for
scattering off an isolated nucleon is~\cite{Harvey:2007rd}
\be\label{eq:xs}
\sigma \approx {\alpha g_\omega^4 G_F^2\over 480\pi^6 m_\omega^4} E_\nu^6 
\approx 2.2\times 10^{-41} (E_\nu / 1\,{\rm GeV})^6 (g_\omega/10)^4 \, {\rm cm}^2 \,.
\ee
The photon energy distribution in this approximation is 
\be
{d\sigma\over dE_\gamma} \propto E_\gamma^3 (E_\nu - E_\gamma)^2 \,,
\ee
and the angular distributions is flat, 
\be
{d\sigma\over d\cos\theta_\gamma} \propto {\rm constant} \,. 
\ee
Form factors will suppress the cross section at large momentum exchange, pulling the angular distribution forward.  As an illustration, the photon 
energy and angular distribution for a $700\,{\rm MeV}$ neutrino incident  
on an isolated nucleon, including nuclear recoil and the form factor 
induced by $\omega(780)$ exchange, is depicted in Fig.~\ref{fig:ff}. 
A more detailed analysis of single photon events will be presented
elsewhere~\cite{HHHnew}. 

\vspace{3mm}
\begin{figure}[h]
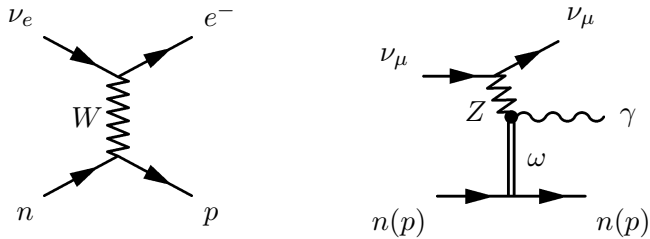

\begin{center}
\begin{fmfchar*}(70,60)
\fmfleftn{l}{2}
\fmfrightn{r}{2}
\fmf{fermion}{l2,w,r2}
\fmf{fermion}{l1,v,r1}
\fmf{zigzag,label=$W$}{w,v}
\fmflabel{$\nu_e$}{l2}
\fmflabel{$e^-$}{r2}
\fmflabel{$n$}{l1}
\fmflabel{$p$}{r1}
\end{fmfchar*}
\hspace{25mm}
\begin{fmfchar*}(70,60)
\fmftopn{t}{5}
\fmfbottomn{b}{5}
\fmfleftn{l}{5}
\fmfrightn{r}{5}
\fmf{phantom}{b3,v,t3}
\fmf{phantom}{l3,v,r3}
\fmffreeze
\fmf{phantom}{l4,w1,w2,w3,w4,r4}
\fmffreeze
\fmf{double,label=$\omega$}{b3,v}
\fmf{fermion}{l4,w2,t4}
\fmf{fermion}{l1,b3,r1}
\fmf{zigzag, label=$Z$, label.side=right}{w2,v}
\fmf{photon}{v,r3}
\fmfdot{v}
\fmflabel{$\nu_\mu$}{l4}
\fmflabel{$\nu_\mu$}{t4}
\fmflabel{$\gamma$}{r3}
\fmflabel{$n(p)$}{l1}
\fmflabel{$n(p)$}{r1}
\end{fmfchar*}
\vspace{2mm}
\end{center}
\caption{\label{fig:misid}
Photon showers from the anomaly-mediated neutral-current 
process (right), can be mistaken for electron showers from 
the charged current process (left). 
On the right, the nucleon can be either neutron or proton. 
}
\end{figure}

In the absence of large coherent enhancements, e.g. for 
scattering on small nuclei, we should ideally 
use relatively large incident neutrino energies, in order to overcome
the mass of $\omega$.  
Also, if it is not possible to distinguish photon showers from 
electron showers, a pure $\nu_\mu$ beam should be used in order to 
avoid a background from charged-current scatters, 
$\nu_e + n \to e^- + p$.  In fact, these requirements have overlap
with experiments looking for $\nu_e$ appearance in a $\nu_\mu$ 
beam.   For example, MiniBooNE~\cite{miniboone} and (in the future) 
T2K~\cite{t2k} have $\nu_\mu$ beams 
with energy spectra of order several hundreds of MeV, but primarily 
$\lesssim 1\,{\rm GeV}$, 
largely within the range of a chiral lagrangian description.   
Single photons that are mistaken for electrons are 
a background to $\nu_e$ appearance searches, as 
depicted in Fig.~\ref{fig:misid}. 
It is interesting that an excess of events observed by MiniBooNE is 
in the same order of magnitude as predicted by (\ref{eq:xs}), and has similar
characteristics to the distributions in Fig.~\ref{fig:ff}.  
Experiments with higher energy neutrinos are also of interest, 
but pass beyond a simple chiral lagrangian description. 

\subsection{Neutrino pair production}

\begin{figure}[h]
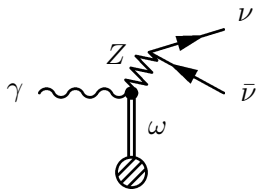

\begin{center}
\begin{fmfchar*}(70,60)
\fmftopn{t}{5}
\fmfbottomn{b}{5}
\fmfleftn{r}{5}
\fmfrightn{l}{5}
\fmf{phantom}{b3,v,t3}
\fmf{phantom}{l3,v,r3}
\fmffreeze
\fmf{phantom}{l4,w1,w2,w3,w4,r4}
\fmffreeze
\fmf{double,label=$\omega$}{b3,v}
\fmf{fermion}{l3,w2,t5}
\fmfblob{4mm}{b3}
\fmf{zigzag, label=$Z$, label.side=right}{w2,v}
\fmf{photon}{v,r3}
\fmflabel{$\nu$}{t5}
\fmflabel{$\bar{\nu}$}{l3}
\fmflabel{$\gamma$}{r3}
\fmfdot{v}
\end{fmfchar*}
\end{center}
\vspace{-2mm}
\caption{\label{fig:nuprod}
Photon conversion into neutrino pairs in the presence of 
baryon number. 
}
\end{figure}
\noindent Similar interactions can give rise to photon conversion into 
neutrino pairs in the presence of baryon number, as depicted 
in Fig.~\ref{fig:nuprod}. 
A nonnegligible contribution to neutron star cooling via this
mechanism was computed in Ref.~\cite{Harvey:2007rd}.  Similar effects will 
occur in the hot and dense environment of a supernova core.  

\subsection{Parity violation} 
Besides neutrino interactions, we can use
the $Z$ to mediate parity violation.  
The interaction (\ref{eq:newint}) will give rise to 
potentially interesting
effects in various parity-violating observables.  These will 
be investigated elsewhere~\cite{HHHnew}. 

\section{Summary}

This report began with a pedagogical derivation 
of the baryon current anomaly in the Standard Model.  The 
counterterm structure in this derivation is interesting because
it requires residual ``pseudo-Chern-Simons'' terms in the action 
when background vector fields are coupled to the quark flavor symmetries.
This exercise is significant for phenomenology 
because the same framework can be used to describe vector meson
interactions in vector dominance approximation.   The resulting
extension of the QCD chiral lagrangian provides a useful 
guide to new effects, such as ``baryon-catalyzed'' neutrino-photon
interactions and parity violation.
Other applications of the formalism that have not been  discussed
here include a description of ``natural parity violating''~%
\footnote{
That is, odd under the parity transformation for which scalars and 
vectors are even, and pseudoscalars and axial-vectors are odd. 
}
QCD vector meson decays, such as $f_1\to\rho\gamma$.
It is also interesting to relate this framework to five-dimensional 
descriptions of QCD~\cite{Sakai:2004cn}, 
both as a means of constraining ``AdS-QCD'' 
models, and potentially using such models to predict undetermined 
constants appearing in the chiral lagrangian. 

\section*{Acknowledgments}
The results reported here are based on Refs.~\cite{Harvey:2007rd,Harvey:2007ca} 
in collaboration
with C.T.~Hill and J.A.~Harvey. 
Research supported by the U.S.~Department of Energy  
grant DE-AC02-76CHO3000.

\end{fmffile}
\end{document}